\documentclass[aps,pra,twocolumn,groupedaddress,noEprint]{revtex4-1}
\usepackage{physics}
\usepackage{graphicx}  %for arXiv
\newcommand{\ii}{{\mathrm{i}}}
\newcommand{\ee}{{\mathrm{e}}}
\begin{document}
\title{
Generation of multiple bound states in the continuum through 
doubly degenerate quasi-guided modes} 

\author{Tetsuyuki Ochiai}
\affiliation{Research Center for Electronic and Optical Materials, National Institute for Materials Science (NIMS), Tsukuba 305-0044, Japan}
\date{\today}

\begin{abstract}
We present a detailed theoretical analysis of a peculiar generation of multiple bound states in the continuum (BICs) in two-dimensional periodic arrays of dielectric spheres. 
They emerge in high-symmetry lattices with the $C_{6v}$ and $C_{4v}$ point groups and involve doubly degenerate quasi-guided modes at the $\Gamma$ point that can couple to external radiation. 
By tuning a system parameter,  the doubly degenerate modes can exhibit accidental BICs at a critical parameter. 
In the vicinity of the critical parameter, the two bands originating from the degenerate mode exhibit multiple off-$\Gamma$ BICs. 
They move and annihilate across the two bands  by changing the  parameter around the critical one.  A ring-like high $Q$ channel pinned with multiple BICs emerges particularly for the $C_{6v}$ case. Across the critical coupling, the total vorticity of the multiple BICs is conserved.   The $\vb*{k}\cdot\vb*{p}$ perturbation theory explains some features of the phenomena reasonably well. 
\end{abstract}

%\keywords{}
\maketitle

\section{Introduction}

Optical bound states in the continuum (BIC) are localized eigenstates embedded in the radiation continuum \cite{hsu2016bound}.  They have  infinite quality factors or, in other words, vanishing decay rates,   although they are inside the light cone. 
Strong light confinement via the infinite quality factor enables us to investigate various applications such as lasing \cite{Kodigala2017}, sensing \cite{Romano2018}, nonlinear optics \cite{Koshelev2020}, and so on, via BICs. 

The BICs are found typically in monolayers of spheres and photonic crystal slabs at the $\Gamma$ point  \cite{Miyazaki1998a,Paddon:Y::61:p2090-2101:2000,Ochiai:S::63:p125107:2001,Fan:J::65:p235112:2002}.  Off-$\Gamma$ BICs are also  found at generic  $\vb*{k}$  points  \cite{Hsu2013b,Jiang2023b}, via the formation of polarization vortices \cite{Zhen2014}. The former BICs are symmetry-protected and irrelevant to physical parameters unless the relevant symmetry is unchanged. The latter are  topologically protected, moving in momentum space.  Therefore, a parameter scan is necessary to find the latter BICs at a given $\vb*{k}$ point other than $\Gamma$. In this sense, the off-$\Gamma$ BICs are sometimes called accidental.

Recently, merging of an at-$\Gamma$
 BIC and off-$\Gamma$ BICs in a non-degenerate isolated band 
attracts much interest as the so-called super BIC \cite{Jin2019,Hwang2021a}.  
This BIC is obtained by tuning system parameters such that the off-$\Gamma$ BICs move toward the $\Gamma$ point, where 
the symmetry-protected BIC exists.  
The super BIC has superb properties as it involves an extreme suppression of the decay rate in a broad region of the momentum space around $\Gamma$. The decay rate  behaves as $|\vb*{k}|^6$, in  striking contrast to the $|\vb*{k}|^2$ behavior of the ordinary symmetry-protected BICs.

Another type of super BICs has been found recently in monolayers of spheres \cite{Kostyukov2022}. From now on, we call it the next-to-super BIC.  It is obtained at the $\Gamma$ point by tuning system parameters and exhibits a $|\vb*{k}|^4$ scaling of the decay rate at a critical parameter. In contrast to the ordinary super BIC, it 
involves doubly-degenerate eigenmodes at the $\Gamma$ point. 
Moreover, off-critical parameters result in a ring-like high $Q$ channel in the triangular lattice system. This high $Q$ channel is either P or S-polarization-like and can be switched by changing the parameter across the critical value.

Once we have a high $Q$ channel, we have a continuous distribution of optical modes with strongly enhanced light-matter interactions. The channel can be selectively excited by the  polarization of the incident light.    
As a result, we can steer the coherent radiation continuously along the ring.  The output light has nearly the same frequency, but the angle changes continuously.   
Such a beam steering paves the way for various applications.

In this paper, we further investigate the next-to-super BICs and  show that multiple off-$\Gamma$ BICs are generated at off-critical parameters nearby. 
We show more complex behaviors than in Ref. \cite{Kostyukov2022} can be observed. 
In addition to the ring-like channel, discrete multiple BICs can be  generated through the next-to-super BICs. In this case, the ring is formed asymmetrically in the parameter space. If we change the parameter across the critical value, the ring disappears, and 
multiple discrete BICs are again generated. Across the critical parameter, the total vorticity of the BICs is conserved. Besides, the appearance of the ring depends on the lattice structure.    

One of the points behind the above phenomena is spatial symmetry. We show that the $\vb*{k}\cdot\vb*{p}$ perturbation theory based on spatial symmetry explains some properties of the phenomena reasonably well.

The ring-like high $Q$ channel can be explained in terms of the multipolar lattice \cite{Sadrieva2019,Gladyshev2022a}.   However, the lattice-structure dependence needs complementary approaches to understand the phenomena. 
The $\vb*{k}\cdot\vb*{p}$ perturbation provides one reasonable scenario as shown in this paper.

The multiple off-$\Gamma$ BICs are also produced by breaking spatial symmetries of  photonic crystal  (PhC) slabs  \cite{Yoda2020,Doiron2022,Wang2023a}. There, degenerate or nondegenerate symmetry-protected BICs at $\Gamma$ turn into multiple off-$\Gamma$ BICs. Since the original symmetry-protected BICs exist irrespective of system parameters, the resulting multiple BICs are deterministic and parameter tuning is  unnecessary.

This paper is organized as follows. In Secs. II and III, we present numerical examples of  the accidental at-$\Gamma$ BICs and resulting multiple off-$\Gamma$ BICs in the triangular and square lattices of dielectric spheres, respectively. In Sec. IV, we analyze these phenomena via the $\vb*{k}\cdot\vb*{p}$ perturbation theory of the Maxwell equation. Finally, in Sec. V, we summarize the results.

\section{Triangular lattice} 

Let us first consider the next-to-super BICs in a monolayer of dielectric spheres arranged in a triangular lattice.  The system has the $C_{6v}$ point group symmetry \cite{inui1996group}.  
Therefore, the eigenmodes are classified according to the irreducible representations of $C_{6v}$.
%and $z$ parity.  
At the $\Gamma$ point, the eigenmodes with the $E_1$ representation of $C_{6v}$ can couple to external radiation. The other eigenmodes of $A_1,A_2,B_1,B_2$ and $E_2$ representations are uncoupled, provided that their eigenfrequencies are below the diffraction threshold. Thus, 
the former eigenmodes generally have finite lifetimes, whereas  the latter eigenmodes have infinite lifetimes and are symmetry-protected BICs.

However, tuning system parameters can cause the former eigenmodes to have  infinite lifetimes. In this way, accidental BICs can occur for the $E_1$ modes.

\begin{figure}
\centering
\includegraphics[width=0.45\textwidth]{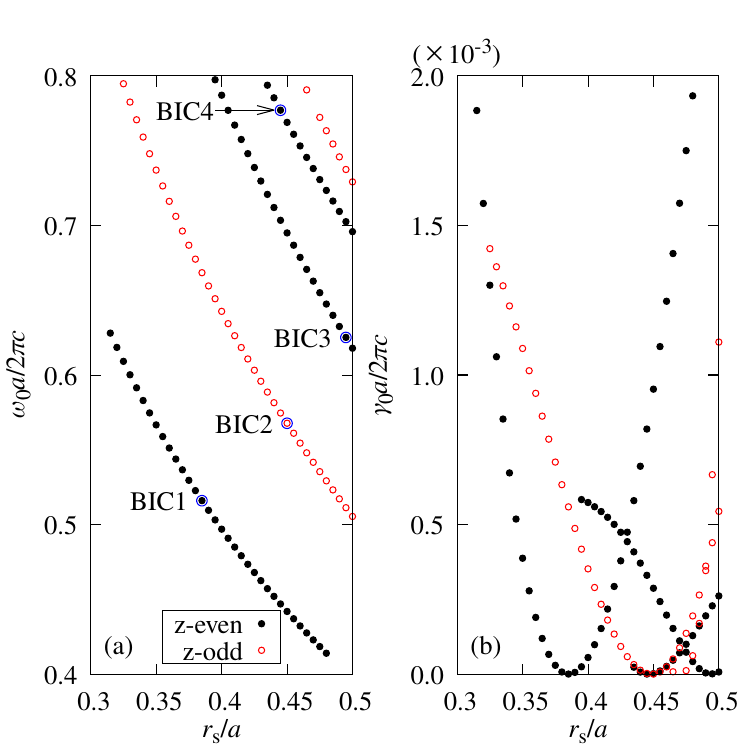}	
\caption{\label{Fig:tr_E1}The resonance angular frequencies $\omega_0$ (a) and widths $\gamma_0$ (b) of the $E_1$ modes at the $\Gamma$ point in the triangular-lattice of dielectric spheres, as a function of sphere radius $r_\mathrm{s}$.  
The modes are classified according to the parity in the $z$ direction, assuming the centers of the spheres are on the $z=0$ plane. The dielectric constant of the spheres is taken to be 12, and that of the background material is 1. The lattice constant is denoted by $a$. Accidental BICs (next-to-super BICs) occur around the modes indicated by the blue circle in (a).        
}	
\end{figure}

Figure \ref{Fig:tr_E1} shows the resonance angular frequency $\omega_k$ and width $\gamma_k$ of the $E_1$ modes at the $\Gamma$ point ($k=0$). 
%% fig.1 (E1 versus rad)%%%%%%%%%%%%%%%%%%%%%%%%%%%%%%%%%%%%%%%
%%%%%%%%%%%%%%%%%%%%%%%%%%%%%%%%%%%%%%%%
They are evaluated with the photonic Korringa-Kohn-Rostoker (KKR) method \cite{Ohtaka1980,MODINOS::141:p575-588:1987,Stefanou1992} together with the curve fitting to the Breit-Wigner formula  \cite{landau2013quantum} of the scattering phase shift $\delta$ as a function of angular frequency $\omega$: 
\begin{align}
\ee^{2\ii\delta}=\ee^{2\ii\delta_k}\qty(\frac{\omega-\omega_k-\ii\gamma_k}{\omega-\omega_k+\ii\gamma_k})^n,
\end{align} 
where $\delta_k$ is the background (frequency-independent) phase shift, and $n$ is the degree of degeneracy of the resonant mode concerned.  
The scattering phase shift is derived from the S-matrix of the monolayer  \cite{Ohtaka:I:Y::70:p035109:2004}. 
At several radii, the eigenmodes exhibit infinite quality factor accidentally.

\begin{figure}
	\centering
	\includegraphics[width=0.45\textwidth]{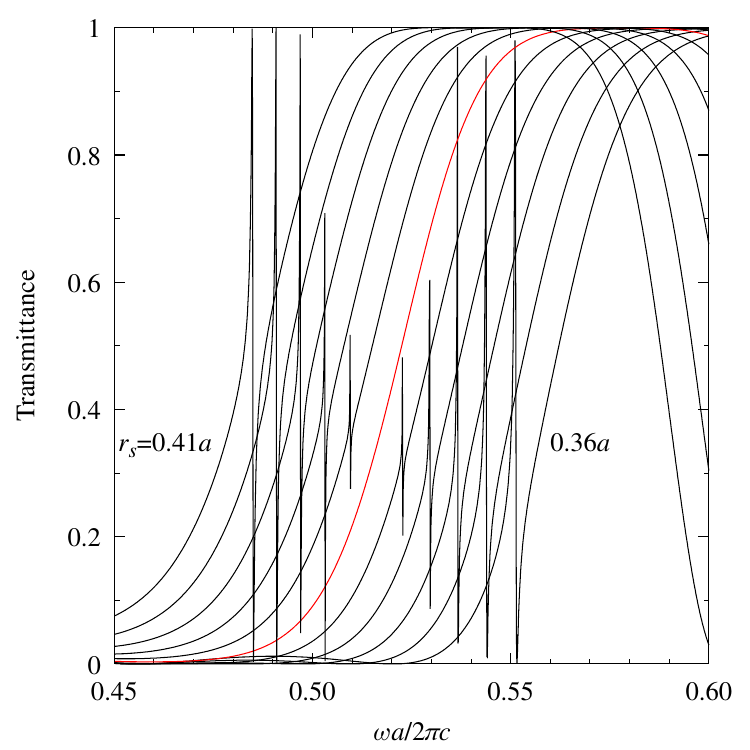}	
	\caption{\label{Fig:tr_G_trans} Specular transmission spectra of the normal incidence in the triangular lattice monolayer of dielectric spheres. The sphere radius varies from $0.36a$ to $0.41a$ with a $0.005a$ step. The other parameters are the same as in Fig. \ref{Fig:tr_E1}. The vanishing resonance signal of the red curve indicates the accidental BIC of BIC1.   
	}	
\end{figure}

Such an accidental BIC can be easily found theoretically and experimentally by monitoring the change in the resonance signal in the  transmission spectrum of the normal incidence.  
For instance, Fig. \ref{Fig:tr_G_trans} shows the change of the transmission spectrum with the sphere radius around the critical one of BIC1 in  Fig. \ref{Fig:tr_E1}.  
%% fig.2 (specular transmittance)%%%%%%%%%%%%%%%%%%%%%%%%%
%%%%%%%%%%%%%%%%%%%%%%%%%%%%%%%%%%%%%%%%
Since the modes other than $E_1$ are symmetry-protected, they do not affect the transmission spectrum. Solely the modes of the $E_1$ representation emerge as an asymmetric resonance signal of the Fano shape \cite{Fano1961} in the spectrum. 
The spectrum changes with the radius, and we can find a narrowing of the resonance width toward the critical radius. The vanishing resonance width corresponds to the accidental BIC.

The above property of the next-to-super BIC presents a marked contrast to the super BIC. In the latter case, the BIC consists of a symmetry-protected BIC with topologically protected BICs, so it is impossible to observe the super BIC via the transmission spectrum of the normal incidence.  It is available only through a detailed analysis of the momentum and parameter dependence of the spectrum.

\begin{figure}
\centering
\includegraphics[width=0.45\textwidth]{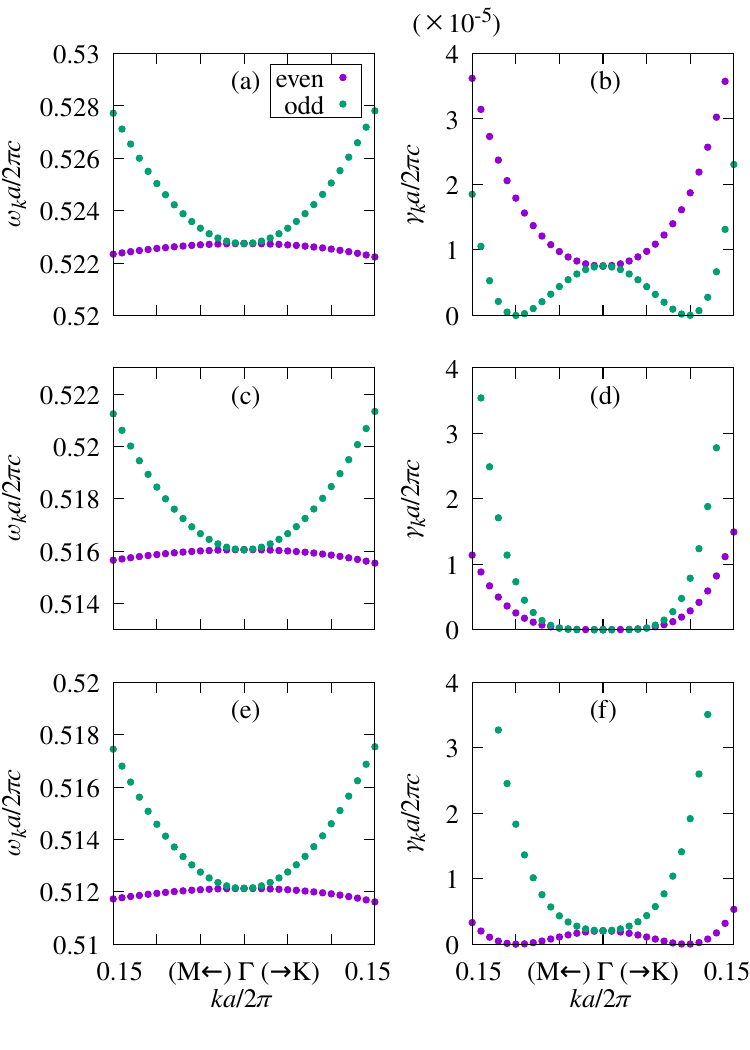}	
\caption{\label{Fig:tr_MGK} Real and imaginary photonic band structures of the $z$-even parity originating from the $E_1$ mode around the critical radius $r_\mathrm{c}$ of BIC1. The momentum is taken along the $\Gamma$M and $\Gamma$K directions. The band structure is classified according to the in-plane parity concerning the $\Gamma$M and $\Gamma$K axes.  The sphere radius is fixed as $r_\mathrm{s}=0.38a$ in (a) and (b), $r_\mathrm{s}=0.385a$ ($\simeq r_\mathrm{c}$) in (c) and (d), and $r_\mathrm{s}=0.388a$  in (e) and (f). The other parameters are the same as in Fig. \ref{Fig:tr_E1}. }	
\end{figure}

Figure \ref{Fig:tr_MGK} shows the real and imaginary photonic band structure around the critical radius of BIC1. 
%% fig.3 (band MGK)%%%%%%%%%%%%%%%%%%%%%%%%%%%%%%%%%
%%%%%%%%%%%%%%%%%%%%%%%%%%%%%%%%%%%%%
Here, the real photonic band structure is referred to as the resonance angular  frequency $\omega_k$, and the imaginary one is to the resonance width $\gamma_k$, as a function of Bloch momentum $k$.    
Although the real band structure does not change so much in its shape, the imaginary band structure  exhibits a clear contrast between the two bands of opposite parities
and among the three values of on- and off-critical radii.  
Below the critical radius [Figs. \ref{Fig:tr_MGK} (a,b)], the odd-parity band exhibits the off-$\Gamma$ BIC at about 
$ka/2\pi=0.1$. 
At the critical radius [Figs. \ref{Fig:tr_MGK}(c,d)], both the bands exhibit 
a flat region of nearly zero values in their imaginary parts. This suppression of the decay rate resembles that in the super BIC. However, now $\gamma_k$ behaves as $k^4$ instead of the $k^6$ behavior of the super BIC.    
Above the critical radius [Figs. \ref{Fig:tr_MGK} (e,f)], the even-parity band  exhibits the off-$\Gamma$ BIC.

\begin{figure}
	\centering
	\includegraphics[width=0.45\textwidth]{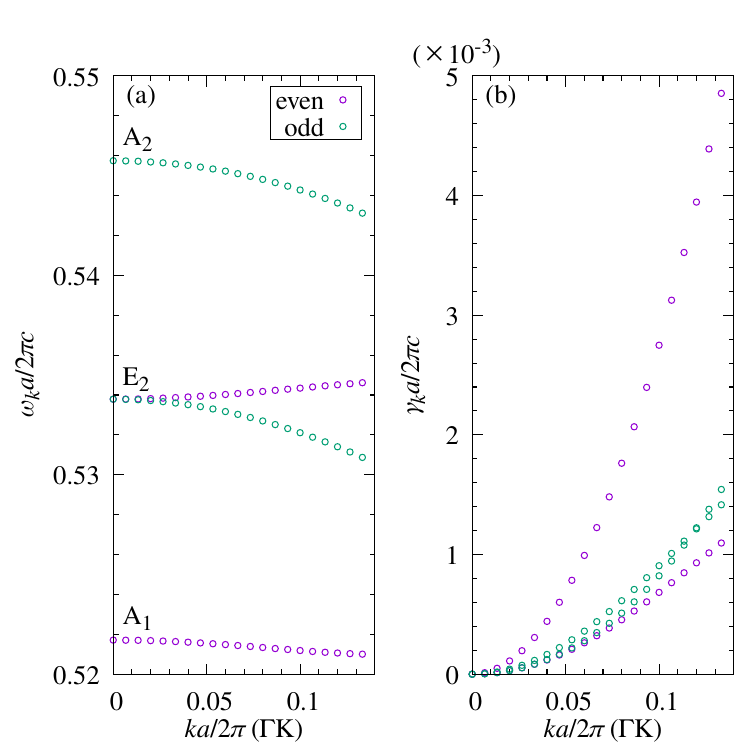}
	\caption{\label{Fig:tr_GK_other} Real and imaginary photonic band structure originating from symmetry-protected BICs at the $\Gamma$ point. 
		The $z$-parity is odd. 
		The momentum is taken along the $\Gamma$K direction.  The sphere radius is fixed as $r_\mathrm{s}=0.38a$.  The other parameters are the same as in Fig. \ref{Fig:tr_E1}. }	
\end{figure}

For comparison, the photonic bands originating from the symmetry-protected BICs at $\Gamma$ exhibit a quick blow-up of the imaginary parts as shown in Fig. \ref{Fig:tr_GK_other}. 
%%%%% Fig.4 (band from symmetry-protected BIC)%%%%%%%%%%%%%%%%%%%%%%%%
%%%%%%%%%%%%%%%%%%%%%%%%%%%%%%%%%%%%%%%%%%%%%%%
Near the $\Gamma$ point, we do not observe the trend of decreasing the imaginary part toward a minimum. 
We should point out that the imaginary part is much larger than  that  from the $E_1$ mode, whose imaginary part is significantly  suppressed.

\begin{figure}
	\centering
	\includegraphics[width=0.45\textwidth]{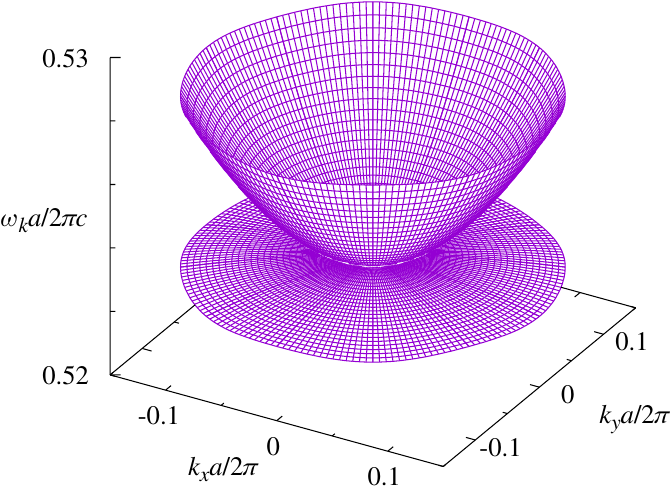}	
	\caption{\label{Fig:tr_bd_all} Real band structure originating from the $E_1$ mode at the $\Gamma$ point. The sphere radius is $0.38a$. The band structure with $|ka/2\pi|\le 0.15$ is plotted. The other parameters are the same as in Fig. \ref{Fig:tr_E1}. }
\end{figure}

By scanning the Bloch momentum in all the directions around the $\Gamma$ point, 
the real band structure consists of two surfaces touched quadratically at the $\Gamma$ point, as shown in Fig. \ref{Fig:tr_bd_all}.  
%%% fig.5 (real band at r=0.38a)%%%%%%%%%%%%%%%%%%%%%%%%%%%%%%
%%%%%%%%%%%%%%%%%%%%%%%%%%%%%%%%%%%%%%%%%%%%%%%%%%
The upper and lower bands are S- and P-polarization-like, respectively.

\begin{figure*}
	\centering
	\includegraphics[width=0.7\textwidth]{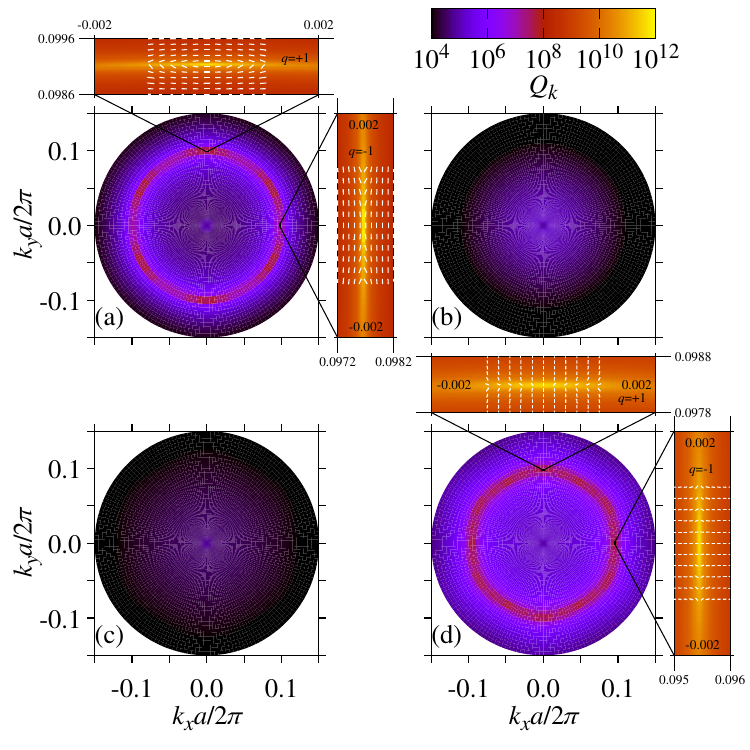}	
	\caption{\label{Fig:tr_ring} $Q$-value map of the two bands  originating from the $E_1$ mode at the $\Gamma$ point near the accidental BIC of BIC1 in Fig. \ref{Fig:tr_E1}. (a,c) The upper (a) and lower (c) bands (in the real band structure) at $r_\mathrm{s}=0.38a (<r_\mathrm{c})$. The real band structure at this parameter is shown in Fig. \ref{Fig:tr_bd_all}. (b,d) The upper (b) and lower (d)  bands at $r_\mathrm{s}=0.388a (>r_\mathrm{c})$.  Insets show the close-up views overlaid by the polarization ellipse map of the photonic band with off-$\Gamma$ BICs.  The vortex charge is denoted by $q$.  }	
\end{figure*}

Figure \ref{Fig:tr_ring} shows the $Q$-value map of the photonic band modes at the off-critical radii.  
%%% fig.6 (Q vlaue map)%%%%%%%%%%%%%%%%%%%%%%%%%%%%%%
%%%%%%%%%%%%%%%%%%%%%%%%%%%%%%%%%%%%%%%%%%%%%%%%%%
It is remarkable that the ring-like high $Q$ channels found in Ref. \cite{Kostyukov2022} are formed.  
The channels lies in the S-polarization-like upper band at $r_\mathrm{s}<r_\mathrm{c}$ and in the P-polarization-like lower band for $r_\mathrm{s}>r_\mathrm{c}$. The other bands (the lower band at $r_\mathrm{s}<r_\mathrm{c}$ and the upper band at $r_\mathrm{s}>r_\mathrm{c}$) do not have such channels. 
These channels indicate that the minimum decay rate $\gamma_k$ at the off-critical radii found in Fig. \ref{Fig:tr_MGK} emerges at nearly the same distance from the $\Gamma$ point regardless of momentum orientation.  
As we move the radius $r_\mathrm{s}$ toward $r_\mathrm{c}$, we can show the ring shrinks to the $\Gamma$ point and across $r_\mathrm{s}=r_\mathrm{c}$, the ring moves from the upper band to the lower band or vice versa.

\begin{figure}
	\centering
	\includegraphics[width=0.25\textwidth]{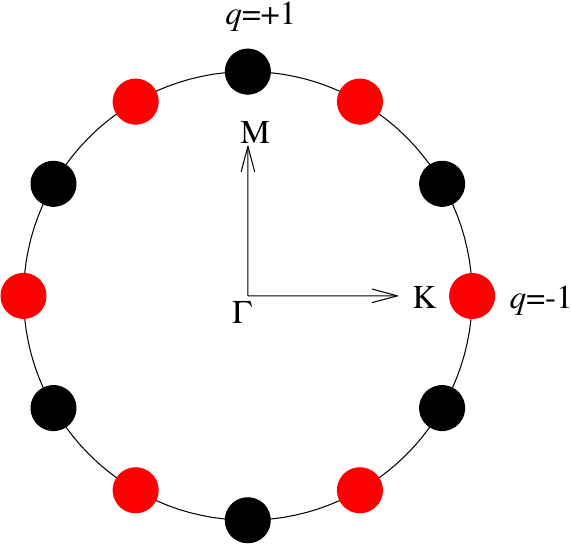}	
	\caption{\label{Fig:tr_BIC1_vortex} Schematic illustration of the vortex charge diagram in the ring. The charge distribution are common between the two rings of Fig. \ref{Fig:tr_ring}.   }	
\end{figure}

Looking closely at the rings, we find twelve off-$\Gamma$ BICs: six on the equivalent $\Gamma$M axes and six on the equivalent $\Gamma$K axes. 
These axes are the mirror axes of the triangular-lattice Brillouin zone. 
The polarization vortices of the BICs are very elongated along the rings.  
The vortex charge diagram is shown in Fig.  \ref{Fig:tr_BIC1_vortex}. 
%%% fig.7 (vorticity in the ring)%%%%%%%%%%%%%%%%%%%%%%%%%%%%%%
%%%%%%%%%%%%%%%%%%%%%%%%%%%%%%%%%%%%%%%%%%%%%%%%%%
The vortex charges are opposite between  $\Gamma$M and $\Gamma$K.  
Off the mirror axes, the $Q$ values in the ring are still very high, of order $10^{-8}$. Thus,  the ring-like high $Q$ channel is of 
quasi-BICs pinned with the true BICs on the mirror axes.

In the above argument, the polarization ellipse in two-dimensional (2D) momentum space is for  
the complex 2D polarization vector $\vb*{e}^+$ defined by 
\begin{align}
&\vb*{e}^+=t_{P\vb*{0}}^+ \hat{\vb*{k}} + t_{s\vb*{0}}^+ \hat{\vb*{k}}_\perp, \\ 
&\vb*{E}^\pm(\vb*{x})=\sum_{\vb*{g}} (t_{P\vb*{g}}^{\pm} \vb*{P}_{\vb*{g}}^\pm + t_{S\vb*{g}}^{\pm} \vb*{S}_{\vb*{g}})\ee^{\ii\vb*{K}_{\vb*{g}}^\pm \cdot\vb*{x}},\label{Eq:Efield} \\
&\vb*{P}_{\vb*{g}}^\pm=\pm \frac{\Gamma_{\vb*{g}}}{q_0}\hat{\vb*{k}}_{\vb*{g}}-\frac{|\vb*{k}_{\vb*{g}}|}{q_0}\hat{z},\quad \vb*{S}_{\vb*{g}}=(\hat{\vb*{k}}_{\vb*{g}})_\perp,\\
&\vb*{K}_{\vb*{g}}^\pm=\vb*{k}_{\vb*{g}}\pm \Gamma_{\vb*{g}}\hat{z},\quad 
\Gamma_{\vb*{g}}=\sqrt{q_0^2-(\vb*{k}_{\vb*{g}})^2}, \\ &\vb*{k}_{\vb*{g}}=\vb*{k}+{\vb*{g}},\quad q_0=\frac{\omega}{c}.
\end{align}	
where $\vb*{E}^\pm(\vb*{x})$ is the electric field of the photonic band mode above (superscript +) and below (-) the monolayer (see Appendix B), $\vb*{k}$ is a 2D Bloch momentum, and $\vb*{g}$ is a 2D reciprocal lattice. 
In the far field, solely the $\vb*{g}=0$ component survives, so that the far-field polarization is determined by $t_{P\vb*{0}}^+ $ and $ t_{S\vb*{0}}^\pm$. 
Besides, the vortex charge $q$ is defined by 
\begin{align}
	&q=\frac{1}{2\pi}\oint \dd\theta_k, 
\end{align}
where $\theta_k$ is the argument of the long axis of the polarization ellipse, 
and the integration contour orbits the vortex core.

\begin{figure}
	\centering
	\includegraphics[width=0.45\textwidth]{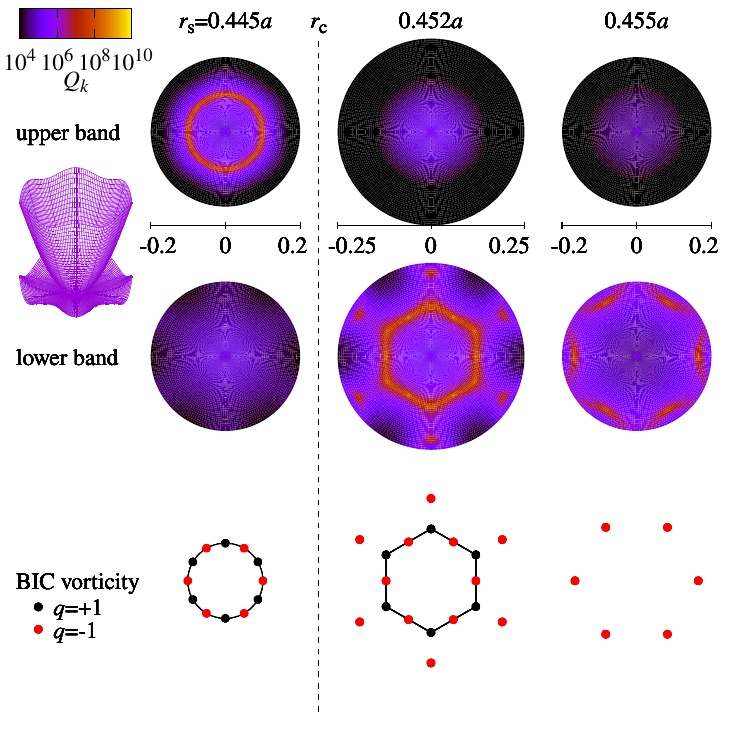}	  
	\caption{\label{Fig:tr_BIC2}  Evolution of multiple BICs around the critical radius $r_\mathrm{c}\simeq 0.45a$ of BIC2 in Fig. \ref{Fig:tr_E1}. The upper two rows show the $Q$ value map of the upper and lower bands originating from the $E_1$ mode at the $\Gamma$ point. 
	The lower row shows the vortex charge diagram of the off-$\Gamma$ BICs.  The inset shows the real band structure around the critical radius. 
	}	
\end{figure}

Around the BIC2 in Fig. \ref{Fig:tr_E1}, we have similar behavior in the ring-like quasi-BICs pinned with multiple true BICs on the mirror axis.  
However, a slightly different behavior than in Fig. \ref{Fig:tr_ring} is observed.   
Figure \ref{Fig:tr_BIC2} shows the evolution of the multiple BICs through the critical radius of BIC2.  
%%%%% Fig.8 (Q value map of BIC2) %%%%%%%%%%%%%%%%%%%%%%%%%%%%%%%%%%%%
%%%%%%%%%%%%%%%%%%%%%%%%%%%%%%%%%%%%%%%%%%%%%%%
Across the critical radius, the ring-like BIC moves from the upper band to the lower band. 
At $r_\mathrm{s}=0.452a(>r_\mathrm{c})$,  there are extra BICs on the $\Gamma$M axes other than the ring-like BICs of the accidental BIC origin.  The latter ring is deformed from a circular shape to a hexagonal shape by a substantial $\vb*{k}\cdot\vb*{p}$ perturbation.  
If we further increase the radius, the extra BICs merge with the true BICs on $\Gamma$M in the ring, and the ring is further deformed.

The BIC3 in Fig. \ref{Fig:tr_E1} is found at $r_\mathrm{s}=0.495a$,  near the close-packing condition of the triangular lattice. Within the photonic KKR method employed in this paper, there is not enough parameter space to enlarge the sphere radius beyond the close-packing condition. We do not consider this BIC here.

\begin{figure}
	\centering
	\includegraphics[width=0.35\textwidth]{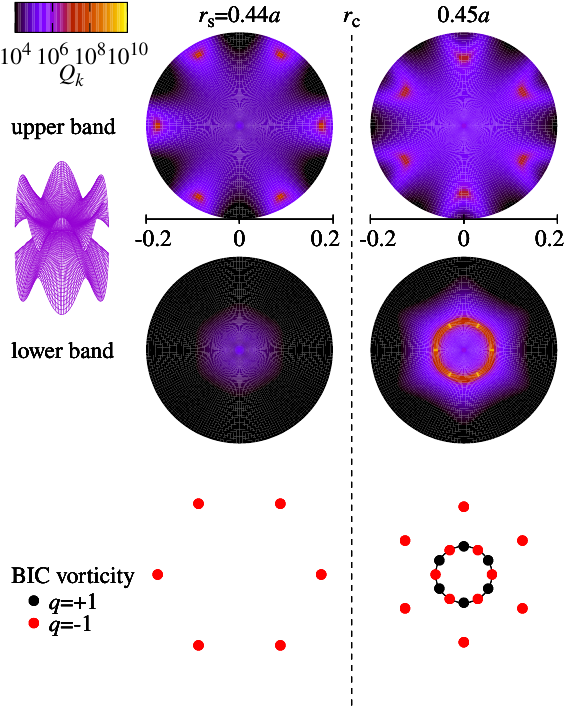}	
	\caption{\label{Fig:tr_BIC4} Evolution of multiple BICs around the critical radius $r_\mathrm{c}\simeq 0.445a$ of BIC4 in Fig. \ref{Fig:tr_E1}. The upper two rows show the $Q$ value map of the upper and lower bands originating from the $E_1$ mode at the $\Gamma$ point.  The lower row shows the vortex charge diagram of the off-$\Gamma$ BICs. The inset shows the real band structure around the critical radius.  
	}	
\end{figure}

The BIC4 has a relatively high resonance frequency,  giving rise to  a more complex multiple BIC distribution  around the critical radius.   
Figure \ref{Fig:tr_BIC4} shows the evolution of the multiple BICs through the critical radius of BIC4. 
%%%%% Fig.9 (Q-value map of BIC4) %%%%%%%%%%%%%%%%%%%%%%%%%%%%%%%%%%%%
%%%%%%%%%%%%%%%%%%%%%%%%%%%%%%%%%%%%%%%%%%%%%%%
Now the ring is formed for the lower band at $r_\mathrm{s}=0.45a (>r_\mathrm{c})$. The upper band exhibits additional BICs on $\Gamma$M. This ring pinned with true BICs on $\Gamma$M and $\Gamma$K, and the additional BICs shrink to the $\Gamma$ point as $r_\mathrm{s}\to r_\mathrm{c}$. Then, at $r_\mathrm{s}=0.44a (<r_\mathrm{c})$, the ring disappears in the lower band, and discrete BICs are generated from the $\Gamma$ point of the upper band. 
The total vorticity is conserved across the critical radius.

\section{Square lattice}

A similar design of the accidental BIC and subsequent multiple BIC generation is available for the square-lattice systems with the $C_{4v}$ point group. A fine-tuning of a system parameter  results in the accidental BIC   
of a doubly degenerate $E$ mode at the $\Gamma$ point.  Above and below the critical parameter, multiple BICs emerge in the two bands originating from the $E$ mode at $\Gamma$. The eigenmodes of $A_1$, $A_2$, $B_1$, and $B_2$ representations of $C_{4v}$ at the $\Gamma$ point are symmetry-protected BICs provided that there are no open diffraction channels other than the specular one.

\begin{figure}
	\centering
	\includegraphics[width=0.45\textwidth]{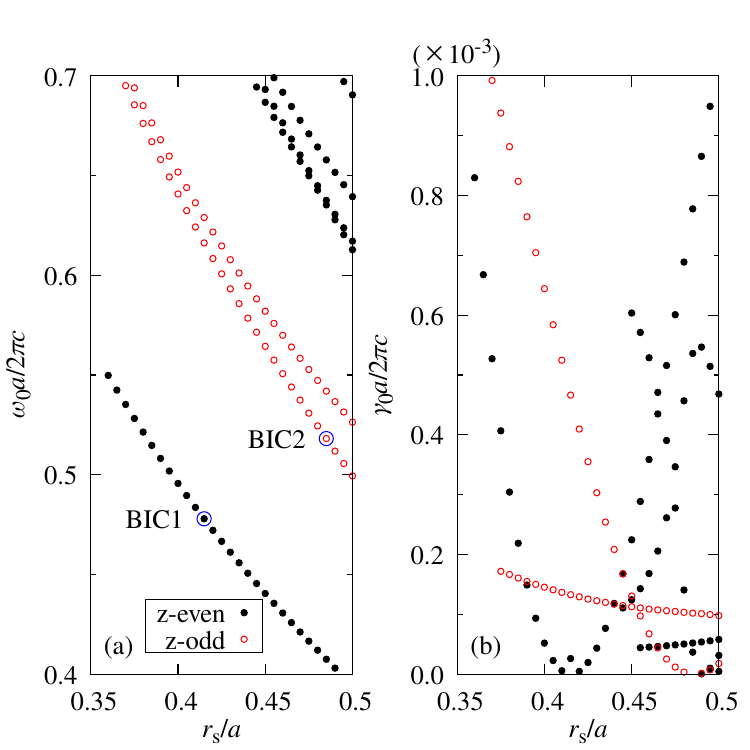}	 
	\caption{\label{Fig:sq_E} Resonance angular frequencies $\omega_0$ (a) and widths $\gamma_0$ (b) of the $E$ modes at the $\Gamma$ point in the square lattice of dielectric spheres, as a function of sphere radius $r_\mathrm{s}$. 	
	The modes are classified according to the parity in the $z$ direction, assuming the centers of the spheres are on the $z=0$ plane. 	The dielectric constant of the spheres is 12, and that of the background material is 1. The lattice constant is denoted by $a$. Accidental BICs occur around the modes indicated by the blue circle in (a). }
\end{figure}

Figure \ref{Fig:sq_E} shows the design of the accidental BIC in the square lattice of identical spheres. 
%%% Fig.8 (E mode vs rad)%%%%%%%%%%%%%%%%%%%%%%%%%%%%%%
%%%%%%%%%%%%%%%%%%%%%%%%%%%%%%%%%%%%%%%%%%%%%%%%%%%%
By changing the sphere radius, we can find two critical radii of the accidental BICs at  $r_\mathrm{s}\simeq 0.415a$ and $0.485a$ for the $z$-even and odd parities, respectively.

\begin{figure}
	\centering
	\includegraphics[width=0.45\textwidth]{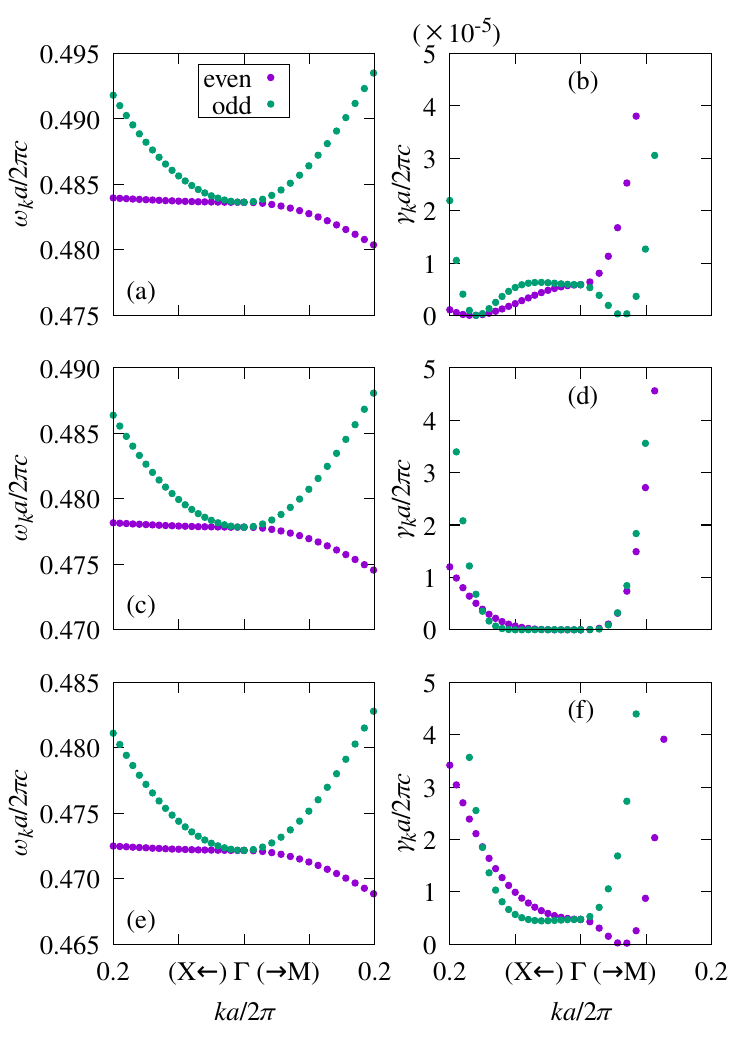}	 
	\caption{\label{Fig:sq_XGM} Real and imaginary photonic band structure  of the $z$-even parity, originating from the $E$ mode at  $\Gamma$ point.  The momentum is scanned in the $\Gamma$X and $\Gamma$M directions.  
	%The coordinate "0.4X(M)" means the point at $\vb*{k}=0.4\vb*{k}_\mathrm{X(M)}$, where  $\vb*{k}_\mathrm{X(M)}$ is the $\vb*{k}$ point at X(M). 
    The photonic bands are further classified according to the parity in these directions.  The sphere radius is $0.41a$ (a,b), $0.415a(\simeq r_\mathrm{c})$ (c,d), and $0.42a$ (e,f). The other parameters are the same as in Fig. \ref{Fig:sq_E}. }
\end{figure}

Considering the parameter regions around the critical radius $r_\mathrm{c}$, we have multiple off-$\Gamma$ BICs. 
Figure \ref{Fig:sq_XGM} shows the real and imaginary photonic band structures of the $z$-even parity around the critical radius of BIC1 
in  Fig. \ref{Fig:sq_E}.  
%%%% fig.11 (band XGM from BIC1)%%%%%%%%%%%%%%%%%%
%%%%%%%%%%%%%%%%%%%%%%%%%%%%%%%%%%%%%%%%%%%%%%%%%%%%%%%
As in Fig. \ref{Fig:tr_MGK}, the real band structure does not change so much, while the imaginary band structure changes remarkably by changing the sphere radius. Concerning the $\Gamma$M direction, a similar trend as in Fig. \ref{Fig:tr_MGK} is observed. Namely, above and below the critical radius, the band that exhibits the off-$\Gamma$ BIC at a nonzero momentum, is interchanged. 
 Below the critical radius [Fig. \ref{Fig:sq_XGM}(a,b)], the odd-parity band exhibits the BIC while above the critical radius [Fig. \ref{Fig:sq_XGM}(e,f)], the even-parity band exhibits the BIC in $\Gamma$M. 
 The suppression of the decay rate at about the critical radius is weaker in $\Gamma$M but is stronger in $\Gamma$X. Such an anisotropy is  manifest also in the real band structure. As for the $\Gamma$X direction, both the even- and odd-parity bands exhibit the off-$\Gamma$ BICs below the critical radius, while the BICs are absent above the critical radius.

\begin{figure}
	\centering
	\includegraphics[width=0.45\textwidth]{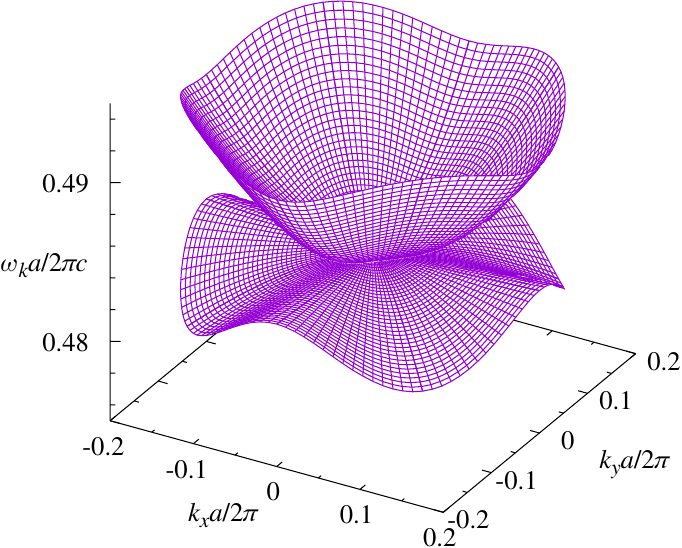} 		 
	\caption{\label{Fig:sq_r041_bd}  Real band structure originating from the $E$ mode at the $\Gamma$ point near the accidental BIC of BIC1 in Fig. \ref{Fig:sq_E}. The sphere radius is $0.41a$. The $z$ parity is even. The band structure with $|ka/2\pi|\le 0.2$ is plotted. The other parameters are the same as in Fig. \ref{Fig:sq_E}. }
\end{figure}

By scanning all the directions around the $\Gamma$ point, the real band structure 
is shown in Fig. \ref{Fig:sq_r041_bd}. 
%%%% Fig. 12 (real band structure of BIC1)%%%%%%%%%%%%%%%%%%%%%%
%%%%%%%%%%%%%%%%%%%%%%%%%%%%%%%%%%%%%%%%%%%%%%%%%
The anisotropy is evident in the band structure.  
The upper band is S-like, and  the lower band is P-like.

\begin{figure*}
	\centering
	\includegraphics[width=0.7\textwidth]{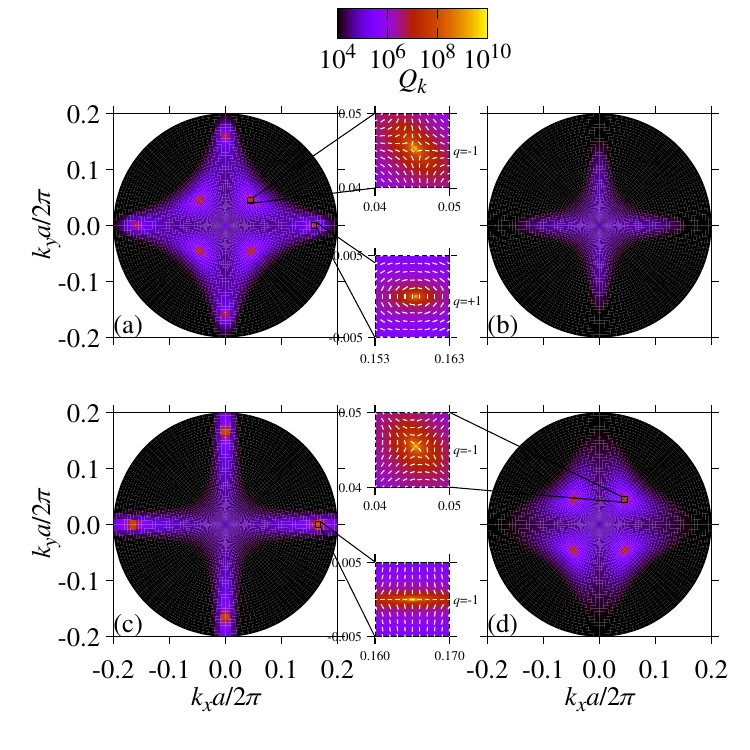} 
	\caption{\label{Fig:sq_BIC1} $Q$-value map of the two bands  originating from the $E$ mode  at the $\Gamma$ point near accidental BIC of BIC1 in Fig. \ref{Fig:sq_E}. (a,c): the upper (a)  and lower (c) bands at $r_\mathrm{s}=0.41a (< r_\mathrm{c})$. (b,d): the upper (b) and lower (d) bands at $r_\mathrm{s}=0.42a (> r_\mathrm{c})$. The insets show the close-up view of the Q-value map overlaid with the polarization ellipse map. }
\end{figure*}

Figure \ref{Fig:sq_BIC1} shows the $Q$-value maps of the two photonic bands originating from the $E$ mode at the off-critical radii of Fig. \ref{Fig:sq_XGM}. 
%%%%Fig. 13 (Q-value map of BIC1)%%%%%%%%%%%%%%%%%%%%%%%%%%%%%%%%%%%%
%%%%%%%%%%%%%%%%%%%%%%%%%%%%%%%%%%%%%%%%%%%%%%%%%
Now, the ring-like BICs found in the triangular-lattice system are absent. A similar pattern as in Fig.  \ref{Fig:sq_BIC1} (a) was observed in Ref. \cite{Kostyukov2022}. 
At $r_\mathrm{s}=0.41a$, the total vorticity vanishes in the upper band. However, the lower band sustains the additional BICs on the $\Gamma$X axes. The net vorticity of the upper and lower bands is thus nonzero. 
All the BICs are discrete and move toward the $\Gamma$ point as $r_\mathrm{s}$ to $r_\mathrm{c}$. At $r_\mathrm{s}=r_\mathrm{c}$, the BICs are collapsed there. Across the critical coupling, multiple BICs are again generated now on the $\Gamma$M axes.  The total vorticity  is conserved across the critical coupling.

\begin{figure}
	\centering
	 \includegraphics[width=0.35\textwidth]{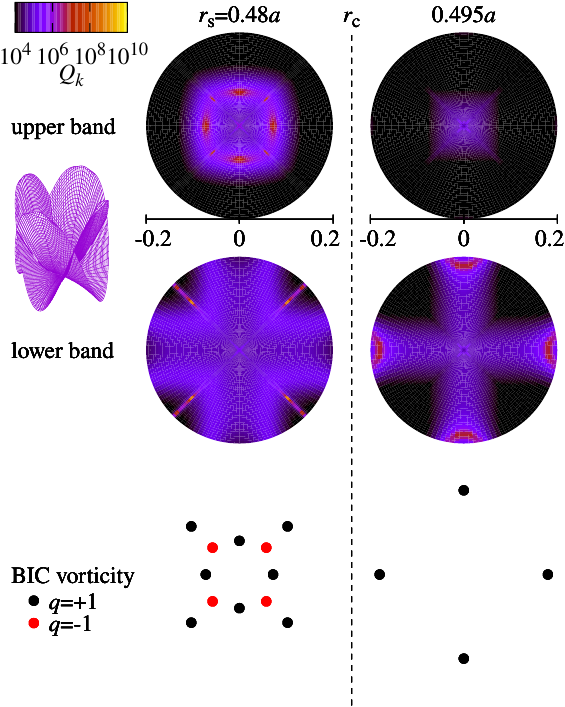}	 
	\caption{\label{Fig:sq_BIC2} $Q$-value map of the photonic bands of the $z$-odd parity, originating from a nearly accidental BIC at the $\Gamma$ point. The inset shows the real band structure (at $r_\mathrm{s}=0.495a$).  }
\end{figure}

As for the BIC2 of Fig. \ref{Fig:sq_E}, the evolution of the multiple BICs across the critical radius is shown in Fig. \ref{Fig:sq_BIC2}. 
%%%%Fig. 14 (Q-value map of BIC2 )%%%%%%%%%%%%%%%%%%%%%%%%%%%%%%%
%%%%%%%%%%%%%%%%%%%%%%%%%%%%%%%%%%%%%%%%%%%%%%%%%
Since the two bands intersect in $\Gamma$M, the Q-value becomes singular there. At $r_\mathrm{s}=0.48a$, the upper band exhibits the eight off-$\Gamma$ BICs. 
They are elongated, and their polarizations are perpendicular and parallel to the $\Gamma$X and $\Gamma$M axes, respectively.  
The lower band holds four off-$\Gamma$ BICs on the $\Gamma$M axes. 
All the BICs moves toward the $\Gamma$ point as $r_\mathrm{s}\to r_\mathrm{c}$. 
After the collapse, the four off-$\Gamma$ BICs are generated as in BIC1.

\section{$\vb*{k}\cdot\vb*{p}$ perturbation theory} 

Let us consider the phenomena obtained in Secs. II and III in the $\vb*{k}\cdot\vb*{p}$ perturbation.  
The eigenmodes in the monolayer are determined by 
\begin{align}
\grad \times \qty(\frac{1}{\varepsilon(\vb*{x})} \grad\times  \vb*{H}(\vb*{x}))=\frac{\omega^2}{c^2} \vb*{H}(\vb*{x}). 
\end{align}
Here, $\vb*{H}$ is the time-harmonic magnetic field with angular frequency $\omega$, and $\varepsilon(\vb*{x})$ is the dielectric function that is periodic in the in-plane coordinate $\vb*{x}_\|$.
The inverse dielectric function  can be expanded  as 
\begin{align}
\frac{1}{\varepsilon(\vb*{x})}=\sum_{\vb*{g}}\ee^{\ii\vb*{g}\cdot\vb*{x}_\|}\eta_{\vb*{g}}(z).
\end{align} 
Besides, the radiation field is also expanded by the plane waves via Bloch's theorem as 
\begin{align}
\vb*{H}(\vb*{x})=\sum_{\vb*{g}}\ee^{\ii(\vb*{k}+\vb*{g})\cdot\vb*{x}_\|}\vb*{h}_{\vb*{g}}(z).  
\end{align}
Accordingly, the equation to be solved becomes  
\begin{align}
&\sum_{\vb*{g}'}\qty(\ii(\vb*{k}+\vb*{g})+\hat{z}\partial_z) \nonumber\\
&\hskip5pt \times\qty[  \eta_{\vb*{g}-\vb*{g}'}(z) \qty(\ii(\vb*{k}+\vb*{g}')+\hat{z}\partial_z)
		\times  \vb*{h}_{\vb*{g}'}(z)]=\frac{\omega^2}{c^2}\vb*{h}_{\vb*{g}}(z),
		\label{Eq:Maxwell} 
\end{align}
which is symbolically expressed as 
\begin{align}
{\cal H}|\psi\rangle={\cal E}|\psi\rangle, \quad {\cal E}=\frac{\omega^2}{c^2}. 	 
\end{align}

Suppose we have a nearly accidental BIC of a doubly degenerate eigenmode at the $\Gamma$ point.    
We perform the  $\vb*{k}\cdot\vb*{p}$ perturbation starting from these states as the zeroth order ones. 
The effective Hamiltonian to be diagonalized is given by \cite{PhysRevB.86.075152}
\begin{align}
&{\cal H}_{ab}^\mathrm{eff}=\langle \phi_{0a}|{\cal H}^{(2)} |\phi_{0b}\rangle \nonumber\\
&\hskip20pt +\sum_{n\ne 0}\frac{1}{\epsilon_0-\epsilon_n}\langle \phi_{0a}|{\cal H}^{(1)} |\phi_{n}\rangle \langle \phi_{n}|{\cal H}^{(1)} |\phi_{0b}\rangle, \label{Eq:effH}\\
&{\cal H}^{(0)} |\phi_{0a}\rangle=\epsilon_0  |\phi_{0a}\rangle, \quad 
{\cal H}^{(0)} |\phi_{n}\rangle=\epsilon_n  |\phi_{n}\rangle,
\end{align}
where ${\cal H}^{(i)}$ is the operator in the left-hand side of Eq. (\ref{Eq:Maxwell}) 
expanded by Bloch momentum $\vb*{k}$ ($i=0,1$ and 2 imply $\vb*{k}=0$, $\vb*{k}$-linear, and $\vb*{k}$-quadratic, respectively).  
The state $|\phi_{0a}\rangle$ ($a=1,2$) represents the doubly degenerate nearly accidental BIC mode at $\Gamma$. 
Here, we assume the degeneracy is not lifted in the first-order perturbation. That is, 
\begin{align}
\langle \phi_{0a}|{\cal H}^{(1)} |\phi_{0b}\rangle=0.
\end{align} 
This property is verified explicitly by the $C_{6v}$ or $C_{4v}$ symmetry. 
More generally, the matrix element $\langle \phi_m|{\cal H}^{(1)} |\phi_n\rangle$ is nonzero only if 
$(\phi_m,\phi_n)$ is attributed to $(E_1,E_2)$, $(E_1,A_1)$, $(E_1,A_2)$, $(E_2,B_1)$, $(E_2,B_2)$, and vice versa for $C_{6v}$, and to $(E,A_1)$, $(E,A_2)$,
 $(E,B_1)$, $(E,B_2)$, and vice versa for $C_{4v}$ (see Appendix A).

The sum in the second term of Eq. \eqref{Eq:effH} generally includes the continuous radiation modes and discrete quasi-guided modes other than the mode concerned, as the intermediate states. The continuous radiation  modes give the decay rate to the original eigenmode via  Fermi's golden rule \cite{Ochiai2001d,Andreani2006}.  However, if we start from the mode of the nearly accidental BIC, the decay rate is strongly suppressed.

The continuum radiation modes belong to the same irreducible representation as the original doubly degenerate mode below the diffraction threshold.  Therefore, the matrix elements vanish between the nearly accidental BIC and radiation modes. That is why the decay rate is strongly suppressed.

In the $C_{6v}$ case, the effective hamiltonian is expressed as 
\begin{align}
&{\cal H}^\mathrm{eff} = \alpha \vb*{k}^2 1 +\beta \qty( (k_x^2-k_y^2)\sigma_3+2k_xk_y\sigma_1 ),\\
&\alpha = \alpha_{E_1E_1}^{(2)} + \sum_{n\in E_2}\frac{|\alpha_{E_1E_2}^{(1)}|^2 }{\epsilon_0-\epsilon_n}\nonumber\\
&\hskip10pt +\frac{1}{2}\sum_{n\in A_1}\frac{|\alpha_{E_1A_1}^{(1)}|^2}{\epsilon_0-\epsilon_n}
+\frac{1}{2}\sum_{n\in A_2}\frac{|\alpha_{E_1A_2}^{(1)}|^2}{\epsilon_0-\epsilon_n},\\
&\beta = \beta_{E_1E_1}^{(2)} \nonumber\\
&\hskip10pt +\frac{1}{2}\sum_{n\in A_1}\frac{|\alpha_{E_1A_1}^{(1)}|^2}{\epsilon_0-\epsilon_n}
-\frac{1}{2}\sum_{n\in A_2}\frac{|\alpha_{E_1A_2}^{(1)}|^2}{\epsilon_0-\epsilon_n}
\end{align}
where $\sigma_i(i=1,2,3)$ is the Pauli matrix, and parameters $\alpha$ and $\beta$ are complex with $\Im[\alpha]\ge 0$. Their imaginary parts come solely from the one in the eigenfrequency of the $E_1$ mode, namely,  $\epsilon_0=(\omega_0-\ii\gamma_0)^2/c^2$, provided that the intermediate modes are lossless ($\mathrm{Im}[\epsilon_n]=0)$.
Therefore, if we start with the $E_1$ mode of the perfect accidental 
BIC ($\gamma_0=0$), $\Im[\alpha]=\Im[\beta]=0$ so that
we still have a vanishing decay rate even at finite $\vb*{k}$, within the second-order $\vb*{k}\cdot\vb*{p}$ perturbation.  This is the case of the extreme suppression of the decay rate around the $\Gamma$ point of the accidental (next-to-super) BIC.   
Since the third-order terms vanish by the inversion symmetry, the imaginary frequencies of the two bands behave $|\vb*{k}|^4$ as pointed out in Ref. \cite{Kostyukov2022}.

By diagonalizing the effective Hamiltonian, the eigenfrequencies of the two bands that stem from the $E_1$ mode at $\Gamma$ are given by 
\begin{align}
&\frac{\omega^2}{c^2}=\frac{(\omega_0-\ii\gamma_0)^2}{c^2} + (\alpha\pm\beta)\vb*{k}^2,  \label{Eq:treffeig}\\
&\vb*{H}_{\vb*{k}}^+(\vb*{x})\propto k_x \vb*{H}_{E_1}^{(01)}(\vb*{x})+k_y\vb*{H}_{E_1}^{(02)}(\vb*{x}),\\
&\vb*{H}_{\vb*{k}}^+(\vb*{x})\propto -k_y \vb*{H}_{E_1}^{(01)}(\vb*{x})+k_x\vb*{H}_{E_1}^{(02)}(\vb*{x}). 
\end{align}
where $(\vb*{H}_{E_1}^{(01)},\vb*{H}_{E_1}^{(02)})$ forms the $E_1$ representation at $\Gamma$. 
If $\gamma_0\ll\omega_0$, the decay rate $\gamma_k$ is evaluated as 
\begin{align}
\gamma_k\simeq\gamma_0-\frac{c^2}{2\omega_0}\Im[\alpha\pm\beta]\vb*{k}^2. 
\end{align}
The radiation field of $\vb*{H}_{\vb*{k}}^{+(-)}$ is P-like (S-like).   
The property of  $\Im[\alpha]>0$ promises a monotonic decreasing of the decay rate $\gamma_k$ with increasing $|\vb*{k}|$ for at least one band, because either $\Im[\alpha+\beta]$ or $\Im[\alpha-\beta]$ is promised to be positive. The behavior of the other band depends on the relative magnitude of $\Im[\alpha]$ and $\Im[\beta]$.  
Across $\mathrm{Im}[\beta]=0$, the band with fast decreasing decay rate with $|\vb*{k}|$ is interchanged.  This condition is satisfied 
at the critical coupling. There, $\gamma_0=0$ and thus $\Im[\beta]=0$.  
Moreover, since the dispersion relation is isotropic, we can have ring-like BICs at the minimum of $\gamma_k$. 
These results explain Fig. \ref{Fig:tr_ring} reasonably well. 

As for Figs. \ref{Fig:tr_BIC2} and \ref{Fig:tr_BIC4}, the spatial anisotropy is very strong, so that the $\vb*{k}\cdot\vb*{p}$ perturbation that predicts the isotropic dispersion is not so efficient.

In the $C_{4v}$ case, the effective Hamiltonian becomes 
\begin{align}
&{\cal H}^\mathrm{eff} = \alpha \vb*{k}^2 1 +\beta_1  (k_x^2-k_y^2)\sigma_3+2\beta_2 k_xk_y\sigma_1,\\
&\alpha = \alpha_{EE}^{(2)} \nonumber\\
&\hskip10pt 
+\frac{1}{2}\sum_{n\in A_1}\frac{|\alpha_{EA_1}^{(1)}|^2}{\epsilon_0-\epsilon_n} +\frac{1}{2}\sum_{n\in B_1}\frac{|\alpha_{EB_1}^{(1)}|^2}{\epsilon_0-\epsilon_n} \nonumber\\
&\hskip10pt 
+\frac{1}{2}\sum_{n\in A_2}\frac{|\alpha_{EA_2}^{(1)}|^2}{\epsilon_0-\epsilon_n}
+\frac{1}{2}\sum_{n\in B_2}\frac{|\alpha_{EB_2}^{(1)}|^2}{\epsilon_0-\epsilon_n}
,\\
&\beta_1 = \beta_{1:EE}^{(2)} \nonumber\\
&\hskip10pt 
+\frac{1}{2}\sum_{n\in A_1}\frac{|\alpha_{EA_1}^{(1)}|^2}{\epsilon_0-\epsilon_n} +\frac{1}{2}\sum_{n\in B_1}\frac{|\alpha_{EB_1}^{(1)}|^2}{\epsilon_0-\epsilon_n} \nonumber\\
&\hskip10pt 
-\frac{1}{2}\sum_{n\in A_2}\frac{|\alpha_{EA_2}^{(1)}|^2}{\epsilon_0-\epsilon_n}
-\frac{1}{2}\sum_{n\in B_2}\frac{|\alpha_{EB_2}^{(1)}|^2}{\epsilon_0-\epsilon_n}
,\\
&\beta_2 = \beta_{2:EE}^{(2)} \nonumber\\
&\hskip10pt 
+\frac{1}{2}\sum_{n\in A_1}\frac{|\alpha_{EA_1}^{(1)}|^2}{\epsilon_0-\epsilon_n} -\frac{1}{2}\sum_{n\in B_1}\frac{|\alpha_{EB_1}^{(1)}|^2}{\epsilon_0-\epsilon_n} \nonumber\\
&\hskip10pt 
-\frac{1}{2}\sum_{n\in A_2}\frac{|\alpha_{EA_2}^{(1)}|^2}{\epsilon_0-\epsilon_n}
+\frac{1}{2}\sum_{n\in B_2}\frac{|\alpha_{EB_2}^{(1)}|^2}{\epsilon_0-\epsilon_n}.
\end{align}
By diagonalizing the effective Hamiltonian, the eigenfrequency becomes 
\begin{align}
	\frac{\omega^2}{c^2}=\frac{(\omega_0-\ii\gamma_0)^2}{c^2} + 
	\alpha\vb*{k}^2 \pm \sqrt{\beta_1^2(k_x^2-k_y^2)^2+4\beta_2^2 k_x^2k_y^2}.
	 \label{Eq:sqeffeig}
\end{align}
Again, $\Im[\alpha]>0$, promising a monotonic decreasing of the imaginary part in Eq. \eqref{Eq:sqeffeig} with increasing $|\vb*{k}|$ for at least one band branch.
Now, Eq. \eqref{Eq:sqeffeig} is not isotropic. Thus, the ring-like high $Q$ channel is not formed. This trend is entirely consistent with the numerical results obtained in Sec. III. 

On the $\Gamma$X axis ($k_y=0$), the two eigenstates become
\begin{align}
&\frac{\omega^2}{c^2}=\frac{(\omega_0-\ii\gamma_0)^2}{c^2} + (\alpha\pm\beta_1)k_x^2,  \label{Eq:sqGXeffeig}\\
&\vb*{H}_{\vb*{k}}^+(\vb*{x})=\vb*{H}_{E}^{(01)}(\vb*{x}), \quad \vb*{H}_{\vb*{k}}^-(\vb*{x})=\vb*{H}_{E}^{(02)}(\vb*{x}),	
\end{align}  
where $(\vb*{H}_{E}^{(01)},\vb*{H}_{E}^{(02)})$ forms the $E$ representation at $\Gamma$. Since the $E$ representation behaves as 
$(x,y)$, $\vb*{H}_{\vb*{k}}^+$ is P-polarized and $\vb*{H}_{\vb*{k}}^-$ is S-polarized.   
One of the eigenmodes is promised to have a decreasing decay rate with 
$|k_x|$, and the two eigenmodes are interchanged at $\Im[\beta_1]=0$ 
via the critical coupling condition $\Im[\gamma_0]=0$. 

On the $\Gamma$M axis ($k_x=k_y$), the eigenstates become
\begin{align}
	&\frac{\omega^2}{c^2}=\frac{(\omega_0-\ii\gamma_0)^2}{c^2} + 2(\alpha\pm \beta_2)k_x^2,  \label{Eq:sqGMeffeig}\\
	&\vb*{H}_{\vb*{k}}^\pm(\vb*{x})\propto \vb*{H}_{E}^{(01)}(\vb*{x})\pm \vb*{H}_{E}^{(02)}(\vb*{x}). 	
\end{align}  
The eigenmodes of superscript $"+"$ is P-polarized, and "-" is  S-polarized. 
One of the two eigenmodes is promised to have a decreasing decay rate with 
$|k_x|$, and the two eigenmodes are interchanged at $\Im[\beta_2]=0$ 
via the critical coupling condition $\Im[\gamma_0]=0$. 
Again, these features are consistent with Figs. \ref{Fig:sq_BIC1} and   \ref{Fig:sq_BIC2}.

The symmetry-protected BICs are also available for $C_{3v}$ point-group systems. There, the eigenmodes of the $A_1$ and $A_2$ representations at the $\Gamma$ point are symmetry-protected, provided there are no open diffraction channels other than the specular one. In this case, it is possible to have accidental at-$\Gamma$ BICs for the eigenmodes of the $E$ representation by tuning system parameters. 

The effective Hamiltonian for the nearly accidental BIC mode is given by 
\begin{align}
 	&{\cal H}^\mathrm{eff} = \alpha \vb*{k}^2 1 +\beta \qty( (k_x^2-k_y^2)\sigma_3+2k_xk_y\sigma_1 ),\\
 	&\alpha = \alpha_{EE}^{(2)} + \sum_{n\in E }{}'\frac{|\alpha_{EE}^{(1)}|^2 }{\epsilon_0-\epsilon_n}\nonumber\\
 	&\hskip10pt +\frac{1}{2}\sum_{n\in A_1}\frac{|\alpha_{EA_1}^{(1)}|^2}{\epsilon_0-\epsilon_n}
 	+\frac{1}{2}\sum_{n\in A_2}\frac{|\alpha_{EA_2}^{(1)}|^2}{\epsilon_0-\epsilon_n},\label{Eq:C3v_alpha}\\
 	&\beta = \beta_{EE}^{(2)} \nonumber\\
 	&\hskip10pt +\frac{1}{2}\sum_{n\in A_1}\frac{|\alpha_{EA_1}^{(1)}|^2}{\epsilon_0-\epsilon_n}
 	-\frac{1}{2}\sum_{n\in A_2}\frac{|\alpha_{EA_2}^{(1)}|^2}{\epsilon_0-\epsilon_n},
\end{align} 
where the prime in the second term of Eq. \eqref{Eq:C3v_alpha} means the unperturbed $E$ mode is excluded in the sum.   
In this case, a significant contribution to the decay rate emerges from the continuous radiation modes of the $E$ representation via Fermi's golden rule. Therefore, suppressing of the decay rate at nonzero $|\vb*{k}|$ near the accidental BIC (at $\Gamma$) does not occur. Consequently, the multiple off-$\Gamma$ BICs are absent in the $\vb*{k}\cdot\vb*{p}$ perturbation theory.  

In the above arguments, we do not rely on the monolayer of spheres presented in Secs. II and III, but on the spatial symmetry of $C_{6v}$, $C_{4v}$, and $C_{3v}$. Accordingly, the multiple BIC generation with the present scenario is available in other photonic membranes with $C_{6v}$ or $C_{4v}$.  We can show PhC slabs with the triangular or square lattice of circular air holes exhibit 
the multiple BIC generation. The ring-like high Q channel is also  formed in the $C_{6v}$ case. 

\section{Summary} 
In summary, we have presented a detailed theoretical analysis of the multiple BIC generation in the monolayers of spheres with $C_{6v}$ or $C_{4v}$ point group. A tuning of system parameters results in an accidental BIC of doubly degenerate eigenmodes at the $\Gamma$ point.   This BIC, obtained at a critical parameter, is the next-to-super BIC with extremely suppressed decay rates around the $\Gamma$ point. Off-critical parameters yield multiple off-$\Gamma$ BICs that are generated from the next-to-super BIC at  the $\Gamma$ point.  

In the $C_{6v}$ point-group system, a ring-like high $Q$ channel pinned with multiple  off-$\Gamma$ BICs on the mirror axes can be formed in the two bands originating from the doubly degenerate $E_1$ mode at the $\Gamma$ point.   
The spatial anisotropy destroys the ring in the $C_{4v}$ point-group system, but multiple BICs are certainly formed. 
Across the critical parameters, these BICs move from the upper band to the lower band or vice versa, conserving the total vorticity of the BICs. 

We also show that these phenomena are owing to the spatial symmetry of $C_{6v}$ or $C_{4v}$ and the $\vb*{k}\cdot\vb*{p}$ perturbation explains the phenomena reasonably well. In addition, the $C_{3v}$ system does not support the multiple BIC generation through a similar design.

\begin{acknowledgments}
This work was supported by JSPS KAKENHI Grant No.  22K03488. 
\end{acknowledgments}

\appendix
\section{Matrix elements in the $\vb*{k}\cdot\vb*{p}$ perturbation}
We summarize various matrix elements relevant to the effective Hamiltonian. The symmetry relation under point-group operations is crucial. 
It is given by 
\begin{align}
{\cal H}_{R_1R_2}^{(i)}(\vb*{k})=D_{R_1}^\dagger(A){\cal H}_{R_1R_2}^{(i)}(A\vb*{k})D_{R_2}(A), \label{Eq:sym}
\end{align}
where the matrix element is the one between the modes of irreducible representations  $R_1$ and $R_2$, and $D_{R}(A)$ is the representation matrix of group element $A$ in the irreducible representation $R$.  
The equation \eqref{Eq:sym} constrains the possible form of the matrix elements.

In the $C_{6v}$ case, 
The nonzero matrix elements in the first-order $\vb*{k}\cdot\vb*{p}$ perturbation are given by 
\begin{align}
&{\cal H}_{E_1E_2}^{(1)}(\vb*{k})=\alpha_{E_1E_2}^{(1)}(k_x\sigma_1+k_y\sigma_3),\\
&{\cal H}_{E_1A_1}^{(1)}(\vb*{k})=\alpha_{E_1A_1}^{(1)}\mqty( k_x\\ k_y),\\
&{\cal H}_{E_1A_2}^{(1)}(\vb*{k})=\alpha_{E_1A_2}^{(1)}\mqty( -k_y\\ k_x),\\
&{\cal H}_{E_2B_1}^{(1)}(\vb*{k})=\alpha_{E_2B_1}^{(1)}\mqty( -k_y\\ k_x),\\
&{\cal H}_{E_2B_2}^{(1)}(\vb*{k})=\alpha_{E_2B_2}^{(1)}\mqty(  k_x\\ k_y).
\end{align}
In the second order, the relevant matrix element becomes 
\begin{align}
{\cal H}_{E_1E_1}^{(2)}(\vb*{k})=\alpha_{E_1E_1}^{(2)}\vb*{k}^2 +\beta_{E_1E_1}^{(2)}((k_x^2-k_y^2)\sigma_3+2k_xk_y\sigma_1), 
\end{align}
where $\alpha_{E_1E_1}^{(2)}$ and $\beta_{E_1E_1}^{(2)}$ are real. 

In the $C_{4v}$ case, 
the nonzero matrix elements in the first-order $\vb*{k}\cdot\vb*{p}$ perturbation  are given by 
\begin{align}
	&{\cal H}_{EA_1}^{(1)}(\vb*{k})=\alpha_{EA_1}^{(1)}\mqty( k_x\\ k_y),\\
	&{\cal H}_{EA_2}^{(1)}(\vb*{k})=\alpha_{EA_2}^{(1)}\mqty( -k_y\\ k_x),\\
	&{\cal H}_{EB_1}^{(1)}(\vb*{k})=\alpha_{EB_1}^{(1)}\mqty( k_x\\ -k_y),\\
	&{\cal H}_{EB_2}^{(1)}(\vb*{k})=\alpha_{EB_2}^{(1)}\mqty(  k_y\\ k_x).
\end{align}
In the second order, the relevant matrix element becomes 
\begin{align}
	{\cal H}_{EE}^{(2)}(\vb*{k})=\alpha_{EE}^{(2)}\vb*{k}^2 1  +\beta_{1:EE}^{(2)}(k_x^2-k_y^2)\sigma_3+ \beta_{2:EE}^{(2)}2k_xk_y\sigma_1, 
\end{align}
where $\alpha_{E_1E_1}^{(2)}$, $\beta_{1:E_1E_1}^{(2)}$, $\beta_{2:E_1E_1}^{(2)}$ are real.

In the $C_{3v}$ case, the nonzero matrix elements in the first-order $\vb*{k}\cdot\vb*{p}$ perturbation  are given by 
\begin{align}
	&{\cal H}_{EE}^{(1)}(\vb*{k})=\alpha_{EE}^{(1)}\mqty( k_x\sigma_1+k_y\sigma_3),\label{Eq:C3v_EE}\\
	&{\cal H}_{EA_1}^{(1)}(\vb*{k})=\alpha_{EA_1}^{(1)}\mqty( k_x\\ k_y),\\
	&{\cal H}_{EA_2}^{(1)}(\vb*{k})=\alpha_{EA_2}^{(1)}\mqty( -k_y\\ k_x). 
\end{align}
The matrix element of Eq. \eqref{Eq:C3v_EE} vanishes by the time-reversal symmetry if the two $E$ modes that form the matrix element are identical.   
In the second order, the relevant matrix element becomes 
\begin{align}
	{\cal H}_{EE}^{(2)}(\vb*{k})=\alpha_{EE}^{(2)}\vb*{k}^2 +\beta_{EE}^{(2)}((k_x^2-k_y^2)\sigma_3+ 2k_xk_y\sigma_1). 
\end{align}

\section{Electric fields of quasi-guided modes} 
We employ the following method for evaluating the polarization ellipse map.  At a true BIC point, the kernel of a secular matrix gives the eigenmode, and its determinant vanishes on the real frequency axis.

In the monolayer of spheres, the secular matrix ${\cal S}$ is given by \cite{Ohtaka1980,Ohtaka1979}
\begin{align}
&{\cal S}_{(L\beta)(L'\beta')}=
\delta_{LL'}\delta_{\beta\beta'}-t_l^\beta G_{LL'}^{\beta\beta'}, \\
&G_{LL'}^{\beta\beta'}=\frac{1}{l(l+1)}\sum_{L_1L_2}[(\vb*{P}^\beta)^\dagger]_{LL_1}G_{L_1L_2} \cdot[\vb*{P}^{\beta'}]_{L_2L'},\\
&G_{LL'}=4\pi\sum_{L''}\ii^{l-l'-l''}\langle L|L''|L'\rangle S_{L''},\\
&S_L=\sum_{\vb*{X}\ne 0}h_l^{(1)}(q_0|\vb*{X}|)Y_L^*(\hat{\vb*{X}})\ee^{\ii\vb*{k}\cdot\vb*{X}}. 
\end{align}
where $L=(l,m)$ ($|m|\le l$) is the angular momentum index, $\beta(=M,N)$ is the index to classify two transverse vector spherical waves, $t_l^\beta$ is  the so-called t-matrix of the sphere, $\vb*{P}^\beta$ is the transformation matrix from scalar spherical waves to vector spherical waves, $\langle L|L''|L'\rangle$ is the Clebsh-Gordan coefficient, $h_l^{(1)}$ is the spherical hankel function of the first kind, $Y_L$ is the spherical harmonics, and $\vb*{X}$ is the 2D real-lattice vector.   
The structure constant $S_L$ is efficiently calculated via the Ewald method \cite{Kambe1967}.  
Through $\psi_L^\beta = \mathrm{Ker}[{\cal S}]$, the electric field outside the monolayer is expressed as 
\begin{align}
&\vb*{E}(\vb*{x})=\sum_{L\vb*{X}}h_l^{(1)}(q_0|\vb*{x}-\vb*{X}|)Y_L(\widehat{\vb*{x}-\vb*{X}})\vb*{V}_L \ee^{\ii \vb*{k}\cdot\vb*{X}},\\
& \vb*{V}_L=\sum_{L'\beta}\vb*{P}_{LL'}^\beta\psi_{L'}^\beta.   	
\end{align}	
Its plane-wave-expansion coefficient $t_{\sigma\vb*{g}}^\pm$ ($\sigma=P,S$) becomes 
\begin{align}
t_{\sigma \vb*{g}}^\pm=	\frac{2\pi}{q_0\Gamma_{\vb*{g}}A_\mathrm{UC}}\sum_{L} (-\ii)^l Y_L(\hat{\vb*{K}}_{\vb*{g}}^\pm)\vb*{\sigma}_{\vb*{g}}^\pm\cdot\vb*{V}_{L},  \label{Eq:SWEtoPWEcoef}
\end{align}
where $A_\mathrm{UC}$ is the area of the unit cell.

Off the BIC point, the determinant is nonzero on the real frequency axis but can be zero in the complex frequency plane. This zero should be 
the resonance pole of the S-matrix, namely, $\Omega_k=\omega_k-\ii\gamma_k$. 
We assume that  the right eigenstate of $\cal{S}$ with the least absolute eigenvalue can be approximated as the kernel on the real frequency axis near the zero. Then, we can obtain the plane-wave  coefficients $t_{\sigma \vb*{g}}^\pm(\omega)$ through Eq.  \eqref{Eq:SWEtoPWEcoef}. 
We further assume that it can be Taylor expanded around $\omega=\Omega_k$, namely, 
\begin{align}
t_{\sigma\vb*{g}}^\pm(\omega)\simeq t_{\sigma\vb*{g}}^\pm(\Omega_k)+(\omega-\Omega_k)C_{\sigma\vb*{g}}^\pm \quad (\sigma=P,S). 
\end{align} 
Then, from the two points $\omega_a$ and $\omega_b$ on the real frequency axis, we can estimate $t_{\sigma\vb*{g}}^\pm(\Omega_k)$ as 
\begin{align}
t_{\sigma\vb*{g}}^\pm(\Omega_k)=\frac{(\omega_a-\Omega_k)t_{\sigma\vb*{g}}^\pm(\omega_b)-(\omega_b-\Omega_k)t_{\sigma\vb*{g}}^\pm(\omega_a)}{\omega_a-\omega_b}.
\end{align} 
We confirm that the resulting $t_{\sigma\vb*{0}}^\pm(\Omega_k)$ vanishes at the BIC points.

%\bibliography{../../Database/library,../../Database/library_add}


\end{document}